\documentclass[twocolumn,amsmath,amssymb,floatfix]{revtex4}

\usepackage{bm}
\usepackage{amsmath}
\usepackage{epsf}
\newcommand{\beq}{\begin{equation}}
\newcommand{\eeq}{\end{equation}} 
\newcommand{\beqa}{\begin{eqnarray}}
\newcommand{\eeqa}{\end{eqnarray}}

\newcommand{\etal}{{\it et al. }}

\newcommand{\fsky}{f_{\rm sky}}

\def\lsim{\lesssim}

\def\ol{\Omega_{\rm de}}
\def\om{\Omega_{\rm m}}
\def\hth{\hat{\phi}}
\def\msun{\,M_\odot}%\def\kms{\ {\rm km\,s^{-1}}}%\def\hmpc{h^{-1}\,{\rm Mpc}}

\newcommand{\ApJL}{Astrophys. J Lett.}
\newcommand{\ApJ}{Astrophys. J}
\newcommand{\PRL}{Phys. Rev. Lett.}
\newcommand{\PRD}{Phys. Rev. D}
\newcommand{\MNRAS}{Mon. Not. Roy. Astr. Soc.}
\newcommand{\ARAA}{Ann. Rev. Astron. Astrophys.}
\newcommand{\AsAs}{Astron. Astrophys.}
\newcommand{\amp}{\& }
\newcommand{\aut}[2]{{#2.\ #1,}}
\newcommand{\laut}[2]{{#2.\ #1,}}
\newcommand{\refs}[6]{#2, {#3},  {#4} (#5).}
\newcommand{\mrefs}[6]{#2, {#3},  {#4} (#5);}
\newcommand{\urefs}[5]{#2, #3, #4 (#5).}
\newcommand{\murefs}[5]{#2, #3, #4 (#5);}
\newcommand{\mybib}[2]{\bibitem{#2}}
\begin{document}

\title{Cross-correlation Tomography: Measuring Dark Energy 
Evolution with Weak Lensing}
\author{Bhuvnesh Jain$^1$ and Andy Taylor$^{2}$}
\affiliation{
{}$^1$ Department of Physics and Astronomy, University of Pennsylvania, 
Philadelphia, PA 19104, U.S.A.\\
{}$^2$ Institute for Astronomy, Royal Observatory,
Blackford Hill, Edinburgh, EH9 3HJ, U.~K.
}

\begin{abstract}
A cross-correlation technique of lensing tomography is presented to measure the
evolution of dark energy in the universe. The variation of the weak lensing
shear with redshift around massive foreground objects like bright galaxies
and clusters depends solely on the angular diameter distances. Use of 
the massive foreground halos allow us to compare 
relatively high, linear shear amplitudes in the same part of the sky, thus 
largely eliminating the
dominant source of systematic error in cosmological weak lensing
measurements. The statistic we use does not rely on knowledge of the 
foreground mass distribution and is only shot-noise limited. 
We estimate the constraints that deep lensing surveys with photometric
redshifts can provide on the dark energy density $\ol$, the equation of 
state parameter $w$ and its redshift derivative $w'$. The marginalized 
accuracies on $w$ and $w'$ are: $\sigma(w)\simeq 0.02\fsky^{-1/2}$ and 
$\sigma(w') \simeq 0.05\fsky^{-1/2}$, where
$\fsky$ is the fraction of sky covered by the survey and $\sigma(\ol)=0.03$
is assumed in the marginalization. 
%The strength of the method lies
%in its constraints on the evolution of dark energy: a deep lensing survey 
%that covers 1/10 of the sky would constrain $w'$ to 10\% accuracy. 
Combining our 
cross-correlation method with standard lensing tomography, which has 
complementary degeneracies, will allow for measurement of the dark energy 
parameters with significantly better accuracy. 
\end{abstract}
\maketitle

\section{Introduction}
%{\parindent0pt\it Introduction.}
Gravitational lensing provides us with the most direct 
method for probing the distribution of matter in the
universe \cite{BarSch01}.
Lensing leads to a shear distortion of background galaxy images, 
or a change in the surface number density of background galaxies due to
magnification. The measurement of the mass distribution in clusters 
of galaxies using the lensing shear \cite{tysoncluster,ks} and
magnification \cite{tyson88,btp,fort,taylor98} are now well established
techniques.  On larger scales detections of the cosmic
shear signal \cite{weakdet} show that the cosmological matter distribution 
can also be probed this way. 

The variation of the lensing signal for
background galaxies at different redshifts probes the projected lensing mass
with different redshift weights in a way that depends on cosmology
\cite{Jain97,Kaiser98,Seljak98,Hu99}. Hu \cite{Hu99, Hu02a, Hu02b} has
developed techniques for using the shear power spectrum for background
galaxies with photometric redshift information to constrain cosmological
parameters, in particular the nature and evolution of dark energy. 
His and other recent studies \cite{Huterer02,Kev02,Heavens03,Refregier03,
Knox03,Linder03} have forecast the accuracy with which these parameters 
can be obtained from future weak lensing surveys, while 
\cite{Taylor01,HuKeeton02} have developed methods to use tomography for 
3-D mass reconstruction. Observationally weak lensing tomography has been 
applied to a galaxy cluster \cite{Wittman01}, but further progress awaits 
multi-color imaging surveys that can obtain photometric redshifts of 
background galaxies. 

The lensing shear (or magnification bias) can be used to 
cross-correlate large foreground galaxies (associated with the lensing
mass) with background galaxies which are lensed.  In this paper we will
use such cross-correlations as an alternative way of doing lensing tomography. 
We use a particularly simple cross-correlation statistic: 
the average tangential shear around massive foreground halos 
associated with galaxy groups and galaxy clusters. 
We show that cross-correlation tomography measures
ratios of angular diameter distances over a range of redshifts. 
The distances are given by integrals of the expansion rate, which in
turn depends on the equation of state of the dark energy. Thus 
it can be used to constrain the evolution of dark energy. 
We estimate the accuracy with which dark energy parameters
can be measured from future lensing surveys. 

\section{Formalism}
%{\parindent0pt\it Formalism.}
We work with the metric 
\begin{equation}
ds^2\, =\, a^2\left[-(1+2\phi)d\tau^2\,+\, (1-2\phi)
    (d\chi^2+r^2d\Omega^2)\right],
    \label{metric}
\end{equation}
where we have used the comoving coordinate $\chi$ and
$a(\tau)=(1+z)^{-1}$ is the scale factor as a function of of conformal
time $\tau$. We adopt units such that $c=1$. The comoving angular
diameter distance $r(\chi)$ depends on the curvature: we assume
a spatially flat universe so that $r(\chi)=\chi$. 
The density parameter $\Omega$ has contributions from mass
density $\om$ or dark energy density $\ol$, so that
$\Omega=\om + \ol$. The dark energy has equation of
state $p=w\rho$, with $w=-1$ corresponding to a cosmological constant. 
The Hubble parameter $H(a)$ is given by 
\begin{equation}
H(a) = H_0\left[\Omega_{m} a^{-3} + 
\Omega_{\rm de} e^{-3  \int_1^a d \ln a' (1+w(a'))}\right]^{1\over 2} ,
\label{hubble}
\end{equation}
where $H_0$ is the Hubble parameter today. The comoving distance $\chi(a)$ is
\begin{equation}
\chi(a) = \int_1^a \frac{d a'}{a'^2 H(a')} ,
\label{chi} 
\end{equation}
The lensing convergence is given by the weighted projection of the mass 
density 
\begin{equation} 
\kappa(\hth)=\frac{3}{2} \om \int_0^{\chi_0} d \chi\ g(\chi)
\frac{\delta(r \hth,\chi)}{a} \; ,
\label{kappa1}
\end{equation}
where the radial weight function $g_{\rm b}(\chi)$ can be expressed in terms
of $r(\chi)$ and the normalized distribution of background galaxies $W_{\rm b}(\chi)$
\begin{equation}
g_{\rm b}(\chi) = r(\chi) \int_\chi^{\chi_0}
{r(\chi' -\chi) \over r(\chi')}W_{\rm b}(\chi')d\chi'\ ,
\label{gchi}
\end{equation}
where $\chi_0$ is the distance to the horizon. For a 
delta-function distribution of background galaxies at $\chi_{\rm b}$, 
this reduces to $g_{\rm b}(\chi)=r(\chi)r(\chi_{\rm b}-\chi)/r(\chi_{\rm b})$. 

We consider the lensing induced cross-correlation between massive foreground
halos, which are traced by galaxies, and the tangential shear with respect 
to the halo center (denoted $\gamma$ in this paper): 
$\omega_\times(\theta)\equiv \left<\delta n_{\rm f}(\hth)\gamma(\hth^\prime)\right>$
where $n_{\rm f}(\hth)$ is the number density of foreground galaxies with mean
redshift $\langle z_{\rm f}\rangle$, observed in the direction $\hth$ in the
sky and $\delta n_{\rm f}(\hth) \equiv (n_{\rm f}(\hth)-{\bar{n}_{\rm f}})/{\bar{n}_{\rm f}}$. 
The angle between directions $\hth$ and $\hth^\prime$ is $\theta$. 
The cross-correlation is given by \cite{Moessner98},\cite{Guzik02}: 
\begin{eqnarray}
\omega_\times (\theta)&=& 6\pi^2 \om \int_0^{\chi_0} d \chi
W_{\rm f}(\chi) \frac{g_{\rm b}(\chi)}{a(\chi)} \nonumber \\
&& \times \int_0^\infty dk\, k\, P_{\rm hm}(\chi, k)\,
J_\mu\left[k r(\chi)\theta\right] \; , 
\label{omegagl}
\end{eqnarray}
where $P_{\rm hm}(\chi, k)$ is the halo-mass cross-power spectrum, 
and $W_{\rm f}$ is the foreground halo redshift distribution. 
The Bessel function
$J_\mu$ has subscript $\mu=2$ for the tangential shear
and $\mu=0$ for the convergence (from the relation
$\gamma(\theta)=-1/2 \ d\, \bar\kappa(\theta)/d\, {\rm ln} \theta$,  
\cite{Guzik02}). The measurement of the mean tangential shear around foreground
galaxies is called galaxy-galaxy lensing. We will consider a generalization
of this to massive halos that span galaxy groups and clusters. 
%The dimensionless angular cross-power spectrum is given by
%\begin{eqnarray}
%2\pi l^2 P(l)= 6 \pi^2 \om l^2 \int_0^{\chi_0} d \chi
% W_{\rm f}(\chi) \frac{g_{b}(\chi, \chi_{b})}{r^2(\chi) a(\chi)} 
%P_{\rm hm}(\chi, k) ,
%\label{powergl}
%\end{eqnarray}
%where $k={l/r(\chi)}$. 

If the foreground sample has a narrow redshift distribution centered
at $\chi=\chi_{\rm f}$, then we
can take $W_{\rm f}$ to be a Dirac-delta function and evaluate the integral
over $\chi$. All terms except $g_{\rm b}(\chi_{\rm f})$ are then 
functions of $\chi_{\rm f}$, the redshift of the lensing mass. 
The coupling of the foreground and background distributions is contained
solely in $g_{\rm b}(\chi_{\rm f})$. Hence 
if we take the ratio of the cross-correlation for two background 
populations with mean redshifts $z_1$ and $z_2$, we get
\begin{equation}
\frac{w_1 (\theta)}{w_2 (\theta)}= 
\frac{g_1(\chi_{\rm f})}{g_2(\chi_{\rm f})} ,
\label{ratio}
\end{equation}
where $w_1, g_1$ denote the values of the functions for the background
population with mean redshift $z_1$. 
In the limit that the background galaxies also 
have a delta-function distribution, this is simply a ratio of distances
\begin{equation}
\frac{w_1 (\theta)}{w_2 (\theta)}= 
\frac{r(\chi_1-\chi_{\rm f})/r(\chi_1)}{r(\chi_2-\chi_{\rm f})/ r(\chi_2)}
\label{ratio2}
\end{equation}
The above equations show that the change in the cross-correlation with
background redshift does not depend on the galaxy-mass 
power spectrum, nor on $\theta$. We can simply use measurements over
a range of $\theta$ to estimate the distance ratio of equation (\ref{ratio2})
for each pair of foreground-background redshifts. The distance ratio in
turn depends on the cosmological parameters $\Omega_{\rm de}$, $w$, and
its evolution $w'$. 
We evaluate next how the cross-correlation defined above can be used 
to constrain the dark energy parameters. 
This approach is complementary to 
standard lensing tomography, where the variation of power spectrum of the shear
with background redshifts is used to measure
an integral of distance and growth factors over redshift. 
With the cross-correlations, there is no dependence on growth 
factor, but a precise dependence on distances (since they are not 
integrated over redshift). 

\section{Results}
\subsection{Signal-to-noise estimate}
%{\parindent0pt\it Results.}
A simple way to estimate the signal-to-noise for the cross-correlation
approach is to regard the foreground galaxies as providing a template
for the shear fields of the background galaxies  (G. Bernstein, private
communication). For a perfect template (i.e. for high density of foreground 
galaxies and no biasing), the errors are solely due to the finite
intrinsic ellipticities of background galaxies. Thus
the fractional error in our measurement of the background shears is simply
\begin{equation}
\frac{\delta \gamma}{\gamma} \sim \frac{\sigma_\epsilon} 
{\sqrt{N_{\rm total}}\ \langle\gamma\rangle_{\rm rms}} .
\label{sn}
\end{equation}
The total number of background galaxies is $N_{\rm total} = n_g A = n_g 
f_{\rm sky} A_{\rm sky}$, where $A$ is the survey area, the lensing 
induced rms shear $\langle\gamma\rangle_{\rm rms} \simeq 0.04$ \cite{Jain97}, 
and the intrinsic ellipticity dispersion $\sigma_\epsilon=0.3$. This gives
\begin{equation}
\frac{\delta \gamma}{\gamma} \sim 0.2 \times 10^{-3} \left(\frac{100}{n_g}
\right)^{1/2} \left(\frac{0.1}{f_{\rm sky}}\right)^{1/2} ,
\label{sn2}
\end{equation}
where the number density $n_g$ has units per square arcminute. 
Thus for the fiducial parameters $\fsky=0.1$ and $n_g=100$, one can 
expect to measure the background shear to 0.1\% accuracy at about 
5-$\sigma$. We find that such a signal corresponds to changes in $w$ of 
a few percent; % (see figure \ref{fig:ratio}); 
hence this is the approximate sensitivity we expect in the absence of 
systematic errors. 

\subsection{Models}
%{\parindent0pt\it Models}
To construct a simple and observationally robust cross-correlation
statistic, we restrict the foreground sample to galaxy clusters and
large galaxy groups. With deep imaging surveys, most of these massive 
objects are expected to be identified out to $z\sim 1$. 
We use the tangential component of the shear for two background samples
at different redshifts, inside apertures
of size $\theta_{\rm ap} \sim 3'$, as our estimator
of the distance ratio of equation (\ref{ratio}). 
The signal from all massive halos in a given redshift bin will be stacked
for each background sample. 
The tangential shear around a foreground halo is 
$\langle\gamma(\theta)\rangle 
= (\bar \Sigma(\theta)-\Sigma(\theta))/\Sigma_{\rm crit}$
where $\Sigma_{\rm crit}$ %=(c^2/4\pi G) (D_s/D_lD_{ls})$  
is the geometrical factor that depends on angular diameter distances. 
To compute the average shear around halos within a mass range, one 
replaces the projected mass density $\Sigma$ by the
projection of the halo-mass correlation function, 
so that the above equation is equivalent to
equation (\ref{omegagl}) in the limit of thin redshift distributions. 
We will assume that the massive foreground halos
at $z<1$ have spectroscopic or accurate photometric redshifts, since they 
generally contain at least one bright galaxy. Each background galaxy 
sample is described by a redshift distribution 
determined using photometric redshifts (which can have relatively 
large statistical errors). 

The average tangential shear around foreground
galaxies and clusters has proven easier to measure than shear-shear
correlations since it is linear in the shear and 
has a distinct signal that can be distinguished from
systematic contributions (for a wide field survey it is unlikely for
a systematic to produce an 
averaged tangential shear around foreground galaxies). Moreover, 
in comparing the signal for two background populations, it is a great
advantage that the shear amplitudes are compared for 
the same set of apertures on the sky: the relative amplitudes are 
insensitive to spatial variations of the point spread function, which 
are the dominant systematic for shear-shear correlations. 

To compute the accuracy with which parameters can be measured, we need
to know the signal and the noise. The signal is given by equation 
(\ref{omegagl}) in which the
halo-mass cross-correlation can be accurately computed using the halo
model of large-scale structure (see \cite{Cooray02} for a review). 
The model specifies $n(m, z)$, 
the comoving number density of halos with mass 
$m$, the bias parameter of halos, and the density
profile of a halo for given mass and redshift. Details of the
model ingredients relevant to this study are given in \cite{Jain03,Takada03}. 

The two key quantities for each redshift bin are the number density
of halos and the mean shear for each halos mass. We choose
the mass range $4\times 10^{13}<m/\msun<10^{15}$ and aperture 
size $\theta_{\rm ap}=2'-3'$ corresponding roughly to the virial radii
of halos over the dominant redshift range. We also set a constant inner
aperture $\theta=0.1'$ to exclude the strong lensing regime and the luminous
parts of the lensing halos. With foreground redshift 
slices of width $\Delta z=0.1$, the halos in each slice cover 
about $1/10$ the survey area. Thus using all the foreground slices 
over $0<z_{\rm f}<1$, close to the entire survey area is covered.

We take the background redshift distribution to be given by
%\begin{equation}
$dN/dz \sim z^a \exp \left[ - \left ( z/z_0 \right)^b \right]$,
%\label{dndz}
%\end{equation}
with $a=2$ and $b=1.5$ and $z_0=1$ giving a mean redshift of $1.5$. 
We normalize it to give a total number of density  $n_g=100$ per square
arcminute. It is split into two background samples, between $z_{\rm f}$ 
and $z=3.0$, with about half the galaxies in each sample. 
The halo model is used to obtain the mean shear for each lens-source
redshift distribution by summing the contributions of halos over the
chosen mass range. While we have used the halo model to compute 
the expected signal-to-noise, the inferred parameter values do not rely on 
knowing the masses or any properties of the halos. 

\subsection{Dark energy parameters}
We perform a $\chi^2$ minimization over the mean shear amplitudes at the
two background distributions for each foreground slice to fit for the dark 
energy parameters. The time dependence of $w$ is parameterized as 
$w=w_0+w_a (1-a)$ \cite{Linder02}. For comparison with other work 
we will compare $w_a$ to $w'$ defined by $w=w_0+w'z$. Since the $w'$ 
parameterization is unsuitable for the large redshift range we 
use, any comparison with it can be made only for a choice of redshift. 
For example, at $z=1$, which is well probed by our method and is of interest
in discriminating dark energy models \cite{Linder02}, 
$w_a = 2 w'$. 
%Following \cite{Hu02b} we choose
%$z_{\rm eff}=0.33$ as the fiducial redshift for $w$, in part because
%$w$ and $w'$ are close to decorrelated with this choice. 
For foreground slices labeled by index $l$ and two background
samples by $1$ and $2$, the $\chi^2$ is given by 
\begin{equation}
\chi^2 = \sum_{l}\left[1 - \frac{R^0(z_l,z_1,z_2)}
{R(z_l,z_1,z_2)}\right]^2         U_{l} , 
\label{chi2}
 \end{equation}
where $R$ is the distance ratio of equation (\ref{ratio}) for given
values of $\ol, w$ and $w_a$, and $R^0$ is the fiducial model with $\ol=0.7, 
w=-1, w_a=0$. The weights $U_{l}$ are given by 
\begin{equation}
U^{-1}_{l} 
= \frac{\sigma_\epsilon^2}{2 n_1 f_l A\ \langle\gamma\rangle^2_{l1}}
+ \frac{\sigma_\epsilon^2}{2 n_2 f_l A\ \langle\gamma\rangle^2_{l2}} ,
\label{weight}
\end{equation}
where $f_l$ is the fraction of the survey area $A$ covered by halo 
apertures in the $l$-th lens slice. The factor of $2$ in the denominator arises
because we are using only one component of the measured ellipticity
whereas $\sigma_\epsilon^2$ denotes the sum of the variances of both
components. 

\begin{figure}[t]
\vspace{7.2cm}
\caption{Contours in the $\Omega_{\rm de}-w$ plane for the fiducial 
lensing survey with $\fsky=0.1$. 
The inner contour assumes no evolution of dark energy, $w_a=0$, 
while the outer contour marginalizes over $w_a$; both employ a
prior on $\ol$ corresponding to $\sigma(\ol)=0.03$ 
(see text). The $68\%$ confidence interval is shown in each of the contours. 
}
\includegraphics{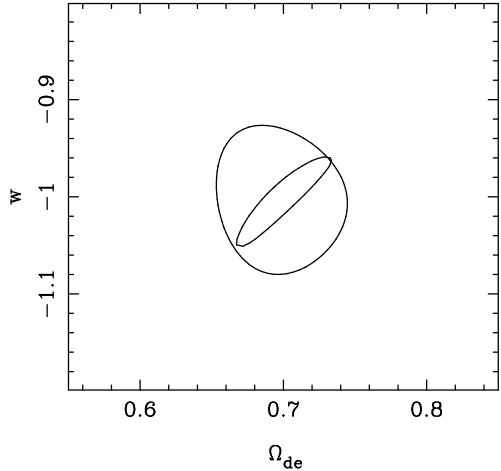}
\label{fig:omegaw}
\end{figure}

\begin{figure}[t]
\vspace{7.2cm}
\caption{Contours in the $w-w'$ plane for the fiducial 
lensing survey with $\fsky=0.1$, as in Figure \ref{fig:omegaw}.  
The inner contour assumes $\ol=0.7$, while the outer contour
marginalizes over $\ol$ with the prior $\sigma(\ol)=0.03$. 
Note that the parameter $w_a = 2\ w'$ at $z=1$ (see discussion in the text). 
}
\includegraphics{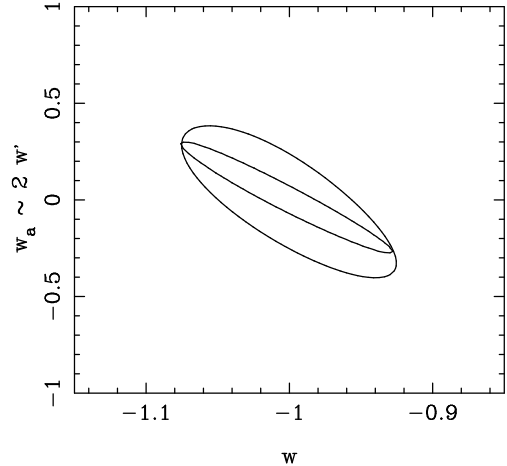}
\label{fig:wdw}
\end{figure}

The results are shown in Figures \ref{fig:omegaw} and \ref{fig:wdw}.
Figure \ref{fig:omegaw} shows the constraints in the $\ol-w$ plane. 
The ellipses show the $68\%$ confidence 
region given by $\Delta\chi^2=2.3$. 
The elongated inner contour is for fixed $w_a=0$, while the outer contour
marginalizes over $w_a$; an external constraint on $\ol$
from the CMB and other probes is assumed, corresponding to $\sigma(\ol)=0.03$.

Figure \ref{fig:wdw} shows
the constraints in the $w-w_a$ plane if $\ol$ is fixed, or
marginalized over with $\sigma(\ol)=0.03$.
The corresponding accuracy on $w$ and $w'$ can be scaled as 
$\fsky^{-1/2}$. For 
$\sigma(\ol)= 0.03$, we obtain
$\sigma(w)\simeq 0.02\fsky^{-1/2}$ and $\sigma(w_a)\simeq 0.1\fsky^{-1/2}$ 
(at $z=1$ this is equivalent to $\sigma(w')\simeq 0.05\fsky^{-1/2}$). 
Note that the value of $\sigma$ on a parameter 
is given by projecting the $\Delta\chi^2=1$
contour on the parameter axis, which is 
smaller than the projection of the inner contour in 
Figure \ref{fig:wdw} with $\Delta\chi^2=2.3$. 
The scaling with $\fsky$ in the parmameter errors comes from the number 
of background galaxies (not sample variance), hence for given $\fsky$ 
varying the depth of the survey scales the errors roughly as $n_g^{-1/2}$. 
The results we have shown are for the fiducial redshift $z=0$. 
A different choice of the fiducial redshift changes the relative 
accuracy on $w$ and $w_a$ somewhat, because the degeneracy direction 
in the three parameters changes. A detailed exploration of different models
of $w(a)$ with finer bins in the background redshift distribution would
be of interest. 

The parameter constraints obtained above are consistent
with the crude signal-to-noise estimate made above in equation (\ref{sn2}). 
This can be seen by using representative numbers for the halos
used in the cross correlation. For $m=10^{14}M_\odot$, 
the number density $n(m)\sim 0.01/{\rm arcmin}^2$ per ln interval in $m$
integrated out to $z=1$, while the mean shear inside $\theta_{\rm
ap}\simeq 2'$ is
$\langle\gamma\rangle \simeq 5\%$ for $z_{\rm f}=0.3$ and $z_{\rm b}=1$
\cite{Takada03}. Thus using halos 
with a mass range of over one decade centered on $m=10^{14}M_\odot$ gives close
to full sky coverage. For relevant scales, the shear profile around the halo 
center is close to the isothermal form: $\gamma\propto1/\theta$. 
The signal-to-noise is then constant per log $\theta$ since the noise 
variance scales as the area ($\propto \theta^2$). 
We will use the simple scheme of 
stacking all halos, and averaging over the mean shear inside apertures,
to estimate the signal. The noise per unit area is 
$\sigma_\epsilon/\sqrt{2 n_g}$. Using the numbers given above 
for $\langle\gamma\rangle$ and
the noise gives a signal-to-noise estimate in agreement with equation 
(\ref{sn2}) as well as the results shown in Figures \ref{fig:omegaw} and 
\ref{fig:wdw}. 

\section{Discussion}
We have presented a method of lensing tomography that uses foreground
halos associated with the lensing mass to measure angular diameter
distances over a range of redshifts, $0\lsim z\lsim 3$. It offers an
observationally robust 
probe of dark energy and its evolution over this redshift range. 
The accuracy achievable with a lensing survey that covers 1/10 of the
sky (with parameters close to that of the survey proposed with the 
LSST telescope) is potentially one of the most promising for proposed surveys
in the coming decade. It is complementary
with constraints from Type 1a supernovae in that the redshift coverage
is broader while the distance factors probed are similar 
\cite{Linder03}. Combining 
cross-correlation tomography with the standard shear power spectrum
tomography improves the constraints significantly, since the contour
ellipses are oriented differently \cite{Hu02b} 
-- this is especially appealing because the two techinques
use the data in different regimes, the linear and strongly nonlinear regime. 
Further it would allow
the constraints on the dark matter, such as on neutrino mass, from shear
power spectrum tomography to be improved by providing independent 
information on the geometry. 

An attractive feature of cross-correlation tomography 
is its observational feasibility. The main practical advantages are:
(i) large shear values ($\sim 1-10\%$) around massive halos are 
used, and (ii) linear shear amplitudes are compared in the {\it same} apertures
on the sky. The statistic is thus largely
insensitive to the primary limiting systematic error in weak lensing surveys:
the variation of the point spread function over the field view (to first 
order the psf variation averages to zero for a linear statistic). 
The imaging requirements are far less stringent than for other cosmological
applications of lensing, such as standard tomography, 
which rely on the measurement of quadratic shear correlations at large angles. 
The ongoing CFH Legacy survey should allow for testing
of the method and parameter constraints that can be checked
with limits from the CMB. The limiting systematic error for most surveys 
is likely come from the photometric redshifts
of the background galaxies. While large statistical errors do not affect
the parameter constraints significantly, systematic errors could bias the 
inferred distance ratios. Calibrating a fair sub-sample of the photometric
redshifts with spectroscopic redshifts, and testing for effects such
as possible correlations between measured ellipticites and the sizes
or surface brightness of galaxies, should allow us to safeguard against
such biases \cite{Bernstein03}. 

We have assumed in this study that the intrinsic ellipticities of 
background galaxies
dominate the statistical errors. Other sources of errors are projection
effects due to matter at different redshifts from the foreground halo and
intrinsic correlations in the ellipticies of background galaxies. Our choice
of averaging the tangential shears around halos should cause 
these errors to not bias the result, but a careful study is needed to quantify
the contribution to the statistical error. A possible bias can be 
introduced if the
foreground redshifts have large errors and the halo selection, 
based on visible galaxies, is perversely correlated with dark energy 
evolution. Statistical accuracy in the foreground redshifts of better
than 0.1 in $z$ should rule out such a bias due to cosmological evolution, 
but again a more detailed study is needed. 

Our implementation of halo-shear cross-correlations is not optimal in that 
we have used only a fraction of the measured shapes for each lens slice, 
and only two bins of the background galaxies (motivated in part by
the finding that this provides most of the information in shear power
spectrum tomography \cite{Hu99}). The parameter accuracy can therefore 
be improved with an optimal scheme in which we use finer binning or 
the actual redshifts
(or photometric redshift probability distributions) of foreground 
and background galaxies rather than binning them. 
It will also be of interest to study different models of $w(z)$ 
rather than the monotonic parameterization used here, or to obtain 
model independent constraints on the expansion history directly
\cite{Tegmark01,Linder02}. A joint analysis with shear-shear 
correlations is needed to quantify the improved precision one can 
expect. 
We have restricted ourselves to the weak lensing shear; additional information
can be obtained by using the strong lensing signal expected in some fraction
of the galaxy and cluster halos (for related studies with lensing arcs
from galaxy clusters, see \cite{Link98,Golse02,Sereno02}) and using the
magnification signal. 

{\it Acknowledgments:} We thank G. Bernstein for many insightful 
conversations. We thank E. Bertschinger, A. Heavens, W. Hu, 
M. Jarvis, E. Linder, U. Pen, R. Sheth and M. Takada for helpful
discussions. 
BJ is supported by NASA grant NAG5-10923 and a Keck 
foundation grant. ANT thanks the PPARC for an Advanced Research Fellowship  
and the University of Pennsylvania for
its hospitality while this work was in development.

\end{document}